\input{aipcheck}

\documentclass[
    ,final            % use final for the camera ready runs
  ,numberedheadings % uncomment this option for numbered sections
  ]
  {aipproc}
\layoutstyle{6x9}
\usepackage{amssymb}
\usepackage{amsmath}
\usepackage{subfigure}
\begin{document}

\title{Fractal Analysis of Weld Defect Patterns Obtained by 
       Radiographic Tests}


\author{Juliano A. Tesser}{
  address={Department of Nuclear Engineering, Federal University of Rio de Janeiro, RJ, Brazil}
}
\author{Ricardo T. Lopes}{
  address={Department of Nuclear Engineering, Federal University of Rio de Janeiro, RJ, Brazil}
}

\author{Andr\'{e} P. Vieira}{
  address={Department of Metallurgical and Materials Engineering, Federal University of Cear\'{a},
           CE, Brazil}
}
\author{Lindberg L. Gon\c{c}alves}{
  address={Department of Metallurgical and Materials Engineering, Federal University of Cear\'{a},
           CE, Brazil}
}

\author{Jo\~{a}o Marcos A. Rebello}{
  address={Department of Metallurgical and Materials Engineering, 
           Federal University of Rio de Janeiro, RJ, Brazil}
  %,altaddress={<author1 address>} % additional visiting address
}

\begin{abstract}
 This paper presents a fractal analysis of radiographic patterns obtained from specimens
with three types of inserted welding defects: lack of fusion, lack of penetration,
and porosity. The study focused on patterns of carbon steel beads from radiographs of the
International Institute of Welding (IIW). The radiographs were scanned using a greyscale
with 256 levels, and the fractal features of the surfaces constructed from the radiographic
images were characterized by means of Hurst, detrended-fluctuation, and minimal-cover
analyses.  A Karhunen-Lo\`{e}ve transformation was then used to classify the curves obtained from the fractal analyses of the various images, and a study of the classification errors was performed.
The obtained results indicate that fractal analyses can be an effective additional tool for
pattern recognition of weld defects in radiographic tests. 
\end{abstract}

\maketitle

%%%%%%%%%%%%%%%%%%%%%%%%%%%%%%%%%%%%%%%%%%%%
%% MAINMATTER
%%%%%%%%%%%%%%%%%%%%%%%%%%%%%%%%%%%%%%%%%%%%

\section{Introduction}

In a recent paper, Silva \emph{et al.} \cite{Silva2005} discussed the characterization of failure
mechanisms that occur in fiberglass-reinforced polymeric matrix composites
when subjected to tensile and flexural loads. The characterization was based
on the analysis of acoustic-emission signals emitted by the composite during
the process of failure. By looking at some fractal properties of the
acoustic emission signals, namely, the fractal indices related to the Hurst
analysis \cite{Hurst1951}, detrended-fluctuation analysis \cite{Peng1994}, 
minimal cover analysis \cite{Dubovikov2004},
and the box-counting dimension analysis \cite{Addison1997}, they were able to
distinguish the different failure modes.

The study presented in this paper aims to characterize, through fractal
analyses, the welding defects present in radiographic images. We focused on patterns
of carbon steel beads from radiographs of the International Institute of Welding (IIW). 
The images were scanned in 8-bit resolution (256 levels of grey), and then processed by using
the software Image Pro Plus 4.0. In order to improve contrast, a median type filter was
used to smooth unpredicted noise. From the scanned images, we built surfaces by associating
the grey level at each pixel with a height variable. The patterns in each surface were then
studied by fractal analyses.

In order to establish the parameters to be calculated, we
first present a brief review of the numerical analysis used in the
treatment of the data. Afterwards, we present and discuss the results
obtained.

\section{Fractal analyses}

\begin{figure}
\begin{centering}
  \includegraphics[height=.4\textheight]{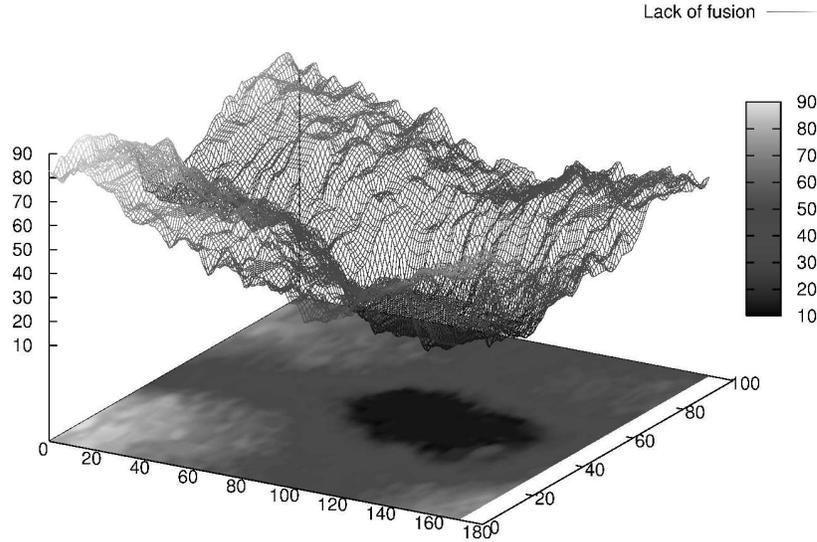}
  \caption{The bottom shows a radiographic image of a defect (lack of fusion), and above it
is the corresponding surface, obtained by converting the greyscale to height variables.}
\end{centering}
\label{fig:1}
\end{figure}

The numerical treatment of the images was performed on data generated
from the 8-bit scanning of the radiographs, corresponding to 256 levels of grey, 
which are translated into a height variable $z_{ij}$. Here,
$(i,j)$ represents the coordinates of a pixel, with $i=1$, $2$, ..., $L_{x}$ and
$j=1$, $2$, ..., $L_{y}$ for an image containing $L_{x}\times L_{y}$ pixels.
This process is illustrated in Fig. \ref{fig:1}

In the fractal analyses we considered extended two-dimensional
versions of the Hurst (or R/S) analysis \cite{Hurst1951}, detrended-fluctuation
analysis (DFA) \cite{Peng1994}, and minimal-cover analysis \cite{Dubovikov2004}.
In order to make the paper self-contained, we will present the details of the
numerical techniques used in the analysis of the surfaces
associated with the images, but first we make a few remarks that are generally valid.

All techniques start by covering the image with a grid composed of square cells containing $\tau\times\tau$ pixels, making sure that the grid is centered in both directions. 
This guarantees that, if $L_{x}$ or $L_{y}$ are not multiples of $\tau$,
only pixels in the periphery of the image are left outside the grid. Each technique
then involves the calculation of the average of some quantity $Q(\tau)$ over all cells, for
different values of $\tau$. In a surface with genuine fractal features,
$Q(\tau)$ should scale as a power of $\tau$ for $\tau\gg 1$,
\[
Q(\tau)\sim\tau^{\eta}.
\]
Fractals of different nature should give rise to different exponents $\eta$, providing
a signature of the fractal. In our case, due to the finite amount of pixels
and the limited resolution of the heights $z_{ij}$, this power-law behavior is hard
to observe. However, as discussed in the final section, we can still use the $\tau$-dependence
of the functions $Q(\tau)$ to characterize the different defects.

\subsection{Hurst analysis}

The rescaled-range (R/S) analysis was introduced by Hurst \cite{Hurst1951} as a tool for 
evaluating the persistency or antipersistency of a time series. The method works by dividing
the series into a series of intervals, and calculating the average ratio of the range (the difference between the maximum and minimum values of the series) to the standard deviation inside
each interval. The size of each interval is then varied.

We define a two-dimensional version of the R/S analysis in the following way. 
Given a $\tau\times\tau$ cell, whose lower left corner is located at pixel $(i_{0},j_{0})$, we calculate $\langle z\rangle_{\tau}$, the average of $z_{ij}$ inside the cell,
\begin{equation}
\langle z\rangle_{\tau}=\frac{1}{\tau^2}{\sum_{(i,j)}}^{\prime}z_{ij}=
\frac{1}{\tau^2}\sum_{i=i_{0}}^{i_{0}+\tau-1}\sum_{j=j_{0}}^{j_{0}+\tau-1}z_{ij},
\end{equation}
where the primed summation runs over all pixels $(i,j)$ inside the cell. We then define an
accumulated deviation from the mean as
\begin{equation}
Z_{ij}=\sum_{k=i_{0}}^{i}\sum_{l=j_{0}}^{j}
\left(z_{kl}-\langle z\rangle_{\tau}\right),
\end{equation}
from which we extract a range,
\begin{equation}
R(\tau)=\max_{\substack{i_{0}\leqslant i\leqslant i_{0}+\tau -1 \\
j_{0}\leqslant j\leqslant j_{0}+\tau -1}}Z_{ij}
-\min_{\substack{i_{0}\leqslant i\leqslant i_{0}+\tau -1 \\
j_{0}\leqslant j\leqslant j_{0}+\tau -1}}Z_{ij}
\end{equation}
and the corresponding standard deviation,
\begin{equation}
S(\tau)=\sqrt{\frac{1}{\tau}{\sum_{(i.j)}}^{\prime}Z_{ij}^2}.
\end{equation}
Finally, we obtain the rescaled range $R(\tau)/S(\tau)$, and take its average over all cells.

In a surface with true fractal features, the rescaled range should satisfy the scaling form
\begin{equation}
\frac{R(\tau)}{S(\tau)}\thicksim \tau^{H},
\end{equation}
where $H$ is the Hurst exponent.

A typical curve obtained from the R/S analysis of the surfaces is shown in Fig. \ref{fig:2}(a).

\begin{ltxfigure}
  \subfigure[RS analysis]{\includegraphics[height=.15\textheight]{rs}}\quad
  \subfigure[DF analysis]{\includegraphics[height=.15\textheight]{df}}\quad
  \subfigure[MC analysis]{\includegraphics[height=.15\textheight]{mc}}
  \caption{Typical curves obtained from the fractal analyses. The quantity $L$ is defined
           as $L=\sqrt{L_{x}L_{y}}$.}
\label{fig:2}
\end{ltxfigure}

\subsection{Detrended-fluctuation analysis}

The detrended-fluctuation analysis (DFA) \cite{Peng1994} aims
to improve the evaluation of correlations in a time series by eliminating trends in the
data. 

Our two-dimensional extension of the method consists initially
in obtaining a new integrated two-dimensional data set ${\tilde{z}}_{ij}$,
\begin{equation}
{\tilde{z}}_{ij}=\sum_{k=1}^{i}\sum_{l=1}^{j}\left(z_{kl}-\langle z\rangle\right),
\end{equation}
where the average $\langle z\rangle$ is taken over all pixels,
\begin{equation}
\langle z\rangle=\frac{1}{L_{x}L_{y}}\sum_{i=1}^{L_{x}}\sum_{j=1}^{L_{y}}z_{ij}.
\end{equation}
After building the grid (with cells of side $\tau$), the integrated data inside
a given cell is fitted by a plane. Then, a detrended variation function
$\Delta_{ij}$ is obtained by subtracting from the integrated data the local trend
as given by the fit. Explicitly, we define
\begin{equation}
\Delta_{ij}={\tilde{z}}_{ij}-h_{ij},
\end{equation}
where $h_{ij}$ is the height associated with pixel $(i,j)$ according to the fit.
Finally, we calculate the root mean-square fluctuation $F(\tau)$ inside a cell as
\begin{equation}
F(\tau)=\sqrt{\frac{1}{\tau^2}{\sum_{(i,j)}}^{\prime}\Delta_{ij}^{2}},
\end{equation}%
and average over all cells. 
For a true fractal surface, $F(\tau)$ should behave as
\begin{equation}
F(\tau)\thicksim\tau^{\alpha},
\end{equation}
where $\alpha $ is the scaling exponent.

A typical curve obtained from the detrented-fluctuation analysis of the surfaces is shown in Fig. \ref{fig:2}(b).

\subsection{Minimal-cover analysis}

This method has been recently introduced \cite{Dubovikov2004}, and,
in its original version, it relates the minimal area necessary to cover a given plane curve, 
at a specified scale, to a power-law behavior. 

In our two-dimensional extension, the method relates the minimal
volume necessary to cover a given surface, at a specified scale. After building
the grid, we can associate with each $\tau\times\tau$ cell, labeled by a variable $k$, a prism of height $A_{k}$, defined as the difference between the maximum and minimum values of $z_{ij}$
inside cell $k$,
\begin{equation}
A_{k}=\max_{\substack{i_{0}\leqslant i\leqslant i_{0}+\tau -1 \\
j_{0}\leqslant j\leqslant j_{0}+\tau -1}}z_{ij}
-\min_{\substack{i_{0}\leqslant i\leqslant i_{0}+\tau -1 \\
j_{0}\leqslant j\leqslant j_{0}+\tau -1}}z_{ij}.
\end{equation}
The minimal volume is then given by
\begin{equation}
V(\tau)=\sum_{k}A_{k}\tau^{2},
\end{equation}
where the summation runs over all cells.

Ideally, in the scaling region ($\tau\gg 1$), $V(\tau)$ should behave as
\begin{equation}
V(\tau )\thicksim \tau ^{3-D_{\mu }},
\end{equation}
where $D_{\mu }$ is the minimal cover dimension, which is equal to $2$ when
the surface presents no fractality.

A typical curve obtained from the minimal-cover analysis of the surfaces is shown in Fig. \ref{fig:2}(c).

\section{Results and discussion}

In order to classify the images, we used a supervised variation of the Karhunen-Lo\`eve (KL) transformation \cite{Webb2002}, and applied it to the set of curves produced by the fractal analyses described in the previous Section. In this sense, the fractal analyses can be seen as a sophisticated preprocessing tool. For each image, we collected the corresponding curves from the 
three fractal analyses, forming a single vector with $M$ components ($M=57$).

We proceeded by first randomly dividing the vectors into a training set (with $N=120$ vectors)
and a test set (with $31$ vectors), performing all the relevant operations (as described 
below), and calculating the confusion tables. Finally, we took averages over $300$ different choices of training and test sets.

Let $\mathbf{x}_{i}$ be the vector corresponding to the $i$th image. The KL transformation
consists of first projecting the training vectors along the eigenvectors of the within-class covariance matrix $\mathbf{S}_{W}$, defined by
\begin{equation}
\mathbf{S}_{W}=\frac{1}{N}\sum_{k=1}^{N_{C}}\sum_{i=1}^{N_{k}}y_{ik}
(\mathbf{x}_{i}-\mathbf{m}_{k})
(\mathbf{x}_{i}-\mathbf{m}_{k})^{T},
\end{equation}
where $N_{C}=4$ is the number of different classes, $N_{k}$ is the number of vectors in class $k$, 
$\mathbf{m}_{k}$ is the average vector of class $k$, and $T$ denotes the transpose of a matrix (in this case, of a column vector). The element $y_{ik}$ is equal to one if $\mathbf{x}_{i}$ belongs
to class $k$, and zero otherwise. 
We also rescale the resulting vectors by a diagonal matrix built 
from the eigenvalues $\lambda_{j}$ of $\mathbf{S}_{W}$. In matrix notation, this operation can be written as
\begin{equation}
\mathbf{X}^{\prime}=\Lambda^{-\frac{1}{2}}\mathbf{U}^{T}\mathbf{X},
\end{equation}
where $\mathbf{X}$ is the matrix whose columns are the training vectors $\mathbf{x}_{i}$,
$\Lambda=\mathrm{diag}(\lambda_{1},\lambda_{2},...)$, and $\mathbf{U}$ is the matrix
whose columns are the eigenvectors of $\mathbf{S}_{W}$. 
This choice of coordinates makes sure that the transformed within-class covariance matrix corresponds to the unit matrix. Finally, in order to compress the class information, 
we project the resulting vectors onto the eigenvectors of the between-class covariance matrix $\mathbf{S}_{B}$,
\begin{equation}
\mathbf{S}_{B}=\sum_{k=1}^{N_{C}}\frac{N_{k}}{N}(\mathbf{m}_{k}-\mathbf{m})
(\mathbf{m}_{k}-\mathbf{m})^{T},
\end{equation}
where $\mathbf{m}$ is the overall average vector.
The full transformation can be written as
\begin{equation}
\mathbf{X}^{\prime\prime}=\mathbf{V}^{T}\Lambda^{-\frac{1}{2}}\mathbf{U}^{T}\mathbf{X},
\end{equation}
where $\mathbf{V}$ is the matrix whose columns are the eigenvectors of $\mathbf{S}_{B}$
(calculated from $\mathbf{X}^{\prime}$).

With $4$ possible classes, the fully-transformed vectors have $4-1=3$ relevant components
\cite{Webb2002}. A vector $\mathbf{x}_i$ is associated with the class whose average vector lies closer to $\mathbf{x}_i$ within the transformed three-dimensional space. 

\begin{table}
\begin{tabular}{ccccc}
\hline
&  LF  &   LP  &  PO & ND  \\ \hline
LF   & 87.4\ (4.8) &  7.5\ (3.6) &  4.9\ (4.3) &  0.2\ (0.9) \\
LP   &  1.5\ (2.1) & 93.8\ (3.4) &  2.1\ (2.6) &  2.5\ (2.4) \\
PO   &  0.2\ (1.0) &  1.4\ (2.7) & 97.0\ (3.8) &  1.4\ (2.6) \\
ND   &  1.3\ (1.2) &  3.7\ (1.8) &  1.6\ (1.1) & 93.4\ (2.3) \\
\hline
\end{tabular}
\caption{Average confusion matrix for the training vectors. The possible classes
are lack of fusion (LF), lack of penetration (LP), porosity (PO) and no defects (ND).
The figures in parenthesis indicate the standard deviations, calculated over $300$
sets. The value in row $i$, column $j$ indicates the percentage of vectors
belonging to class $i$ which were associated with class $j$.}
\label{tab:1}
\end{table}

\begin{table}
\begin{tabular}{ccccc}
\hline
&  LF  &   LP  &  PO & ND  \\ \hline
LF   &  48.1\ (23.6) &  19.8\ (20.7) &  22.8\ (19.9) &   9.2\ (14.7) \\
LP   &  10.7\ (11.9) &  55.1\ (18.8) &  14.3\ (14.0) &  19.9\ (16.1) \\
PO   &  12.2\ (16.3) &  14.7\ (18.6) &  57.6\ (25.0) &  15.5\ (19.1) \\
ND   &   9.4\ (8.7)  &  19.2\ (11.2) &  13.7\ (10.1) &  57.6\ (14.3) \\
\hline
\end{tabular}
\caption{The same as in Table \ref{tab:1}, for the testing vectors.}
\label{tab:2}
\end{table}

After using the training vectors to obtain the full transformation, the classification can
be checked with those same vectors. The resulting average confusion matrix is displayed in
Table \ref{tab:1}. When employed to classify the testing vectors, the KL transformation
produces the confusion matrix shown in Table \ref{tab:2}. From the tables we can see that
the training vectors are associated with the correct class in about $90\%$ of the cases,
whereas the percentage of correct classification for the testing vectors is about $50\%$.
This last result surely lies above the expected rate
produced by random association (which would be $25\%$ in this case).

In summary, we have shown that fractal analyses are a promising tool for 
classifying welding defects in radiographic images. 
We believe that its efficiency can be considerably improved by using a larger sample of images,
as well as by working with 16-bit scans, which would greatly enhance the grey-level resolution.
We are currently working on this direction.

\begin{theacknowledgments}
This work was partially financed by the Brazilian agencies CNPq, FUNCAP and CAPES.
\end{theacknowledgments}



\bibliographystyle{aipproc}   % if natbib is available

\bibliography{qnde2006}

%  \typeout{******************************************}
%  \typeout{** Please run "bibtex \jobname" to optain}
%  \typeout{** the bibliography and then re-run LaTeX}
%  \typeout{** twice to fix the references!}
%  \typeout{******************************************}

\end{document}